\documentclass[a4paper,11pt]{article}
\pdfoutput=1 
\usepackage[T1]{fontenc} 

\makeatletter
\gdef\@fpheader{ }
\gdef\@journal{ }
\makeatother
\RequirePackage{amsmath}
\RequirePackage{amssymb}
\RequirePackage{epsfig}
\RequirePackage{graphicx}
\RequirePackage[numbers,sort&compress]{natbib}
\RequirePackage{color}
\RequirePackage[colorlinks=true
,urlcolor=blue
,anchorcolor=blue
,citecolor=blue
,filecolor=blue
,linkcolor=blue
,menucolor=blue
,pagecolor=blue
,linktocpage=true
,pdfproducer=medialab
,pdfa=true
]{hyperref}

\newif\ifnotoc\notocfalse
\newif\ifemailadd\emailaddfalse
\newif\iftoccontinuous\toccontinuousfalse
\makeatletter
\def\@subheader{\@empty}
\def\@keywords{\@empty}
\def\@abstract{\@empty}
\def\@xtum{\@empty}
\def\@dedicated{\@empty}
\def\@arxivnumber{\@empty}
\def\@collaboration{\@empty}
\def\@collaborationImg{\@empty}
\def\@proceeding{\@empty}
\def\@preprint{\@empty}

\newcommand{\subheader}[1]{\gdef\@subheader{#1}}
\newcommand{\keywords}[1]{\if!\@keywords!\gdef\@keywords{#1}\else%
\PackageWarningNoLine{\jname}{Keywords already defined.\MessageBreak Ignoring last definition.}\fi}
\renewcommand{\abstract}[1]{\gdef\@abstract{#1}}
\newcommand{\dedicated}[1]{\gdef\@dedicated{#1}}
\newcommand{\arxivnumber}[1]{\gdef\@arxivnumber{#1}}
\newcommand{\proceeding}[1]{\gdef\@proceeding{#1}}
\newcommand{\xtumfont}[1]{\textsc{#1}}
\newcommand{\correctionref}[3]{\gdef\@xtum{\xtumfont{#1} \href{#2}{#3}}}
\newcommand\jname{JHEP}
\newcommand\acknowledgments{\section*{Acknowledgments}}

\newcommand\preprint[1]{\gdef\@preprint{\hfill #1}}

\makeatother

\newcommand\note[2][]{%
\if!#1!%
\stepcounter{footnote}\footnotetext{#2}%
\else%
{\renewcommand\thefootnote{#1}%
\footnotetext{#2}}%
\fi}

\makeatletter
\newtoks\auth@toks
\renewcommand{\author}[2][]{%
  \if!#1!%
    \auth@toks=\expandafter{\the\auth@toks#2\ }%
  \else
    \auth@toks=\expandafter{\the\auth@toks#2$^{#1}$\ }%
  \fi
}
\makeatother
\makeatletter
\newtoks\affil@toks\newif\ifaffil\affilfalse
\newcommand{\affiliation}[2][]{%
\affiltrue
  \if!#1!%
    \affil@toks=\expandafter{\the\affil@toks{\item[]#2}}%
  \else
    \affil@toks=\expandafter{\the\affil@toks{\item[$^{#1}$]#2}}%
  \fi
}
\makeatother
\makeatletter
\newtoks\email@toks\newcounter{email@counter}%
\newcommand{\emailAdd}[1]{%
\emailaddtrue%
\ifnum\theemail@counter>0\email@toks=\expandafter{\the\email@toks, \@email{#1}}%
\else\email@toks=\expandafter{\the\email@toks\@email{#1}}%
\fi\stepcounter{email@counter}}
\newcommand{\@email}[1]{\href{mailto:#1}{\tt #1}}
\makeatother

\makeatletter
\newcommand*\collaboration[1]{\gdef\@collaboration{#1}}
\newcommand*\collaborationImg[2][]{\gdef\@collaborationImg{#2}}
\makeatletter
\newcommand\afterLogoSpace{\smallskip}
\newcommand\afterSubheaderSpace{\vskip3pt plus 2pt minus 1pt}
\newcommand\afterProceedingsSpace{\vskip21pt plus0.4fil minus15pt}
\newcommand\afterTitleSpace{\vskip23pt plus0.06fil minus13pt}
\newcommand\afterRuleSpace{\vskip23pt plus0.06fil minus13pt}
\newcommand\afterCollaborationSpace{\vskip3pt plus 2pt minus 1pt}
\newcommand\afterCollaborationImgSpace{\vskip3pt plus 2pt minus 1pt}
\newcommand\afterAuthorSpace{\vskip5pt plus4pt minus4pt}
\newcommand\afterAffiliationSpace{\vskip3pt plus3pt}
\newcommand\afterEmailSpace{\vskip16pt plus9pt minus10pt\filbreak}
\newcommand\afterXtumSpace{\par\bigskip}
\newcommand\afterAbstractSpace{\vskip16pt plus9pt minus13pt}
\newcommand\afterKeywordsSpace{\vskip16pt plus9pt minus13pt}
\newcommand\afterArxivSpace{\vskip3pt plus0.01fil minus10pt}
\newcommand\afterDedicatedSpace{\vskip0pt plus0.01fil}
\newcommand\afterTocSpace{\bigskip\medskip}
\newcommand\afterTocRuleSpace{\bigskip\bigskip}
\newlength{\affiliationsSep}
\newcommand\beforetochook{\pagestyle{myplain}\pagenumbering{roman}}

\DeclareFixedFont\trfont{OT1}{phv}{b}{sc}{11}

\renewcommand\maketitle{
\pagestyle{empty}
\thispagestyle{titlepage}
\setcounter{page}{0}
\noindent{\small\scshape\@fpheader}\@preprint\par

\afterLogoSpace
\if!\@subheader!\else\noindent{\trfont{\@subheader}}\fi
\afterSubheaderSpace
\if!\@proceeding!\else\noindent{\sc\@proceeding}\fi
\afterProceedingsSpace
{\LARGE\flushleft\sffamily\bfseries\@title\par}
\afterTitleSpace
\hrule height 1.5\p@%
\afterRuleSpace
\if!\@collaboration!\else
{\Large\bfseries\sffamily\raggedright\@collaboration}\par
\afterCollaborationSpace
\fi
\if!\@collaborationImg!\else
{\normalsize\bfseries\sffamily\raggedright\@collaborationImg}\par
\afterCollaborationImgSpace
\fi
{\bfseries\raggedright\sffamily\the\auth@toks\par}
\afterAuthorSpace
\ifaffil\begin{list}{}{%
\setlength{\leftmargin}{0.28cm}%
\setlength{\labelsep}{0pt}%
\setlength{\itemsep}{\affiliationsSep}%
\setlength{\topsep}{-\parskip}}
\itshape\small%
\the\affil@toks
\end{list}\fi
\afterAffiliationSpace
\ifemailadd 
\noindent\hspace{0.28cm}\begin{minipage}[l]{.9\textwidth}
\begin{flushleft}
\textit{E-mail:} \the\email@toks
\end{flushleft}
\end{minipage}
\else 
\PackageWarningNoLine{\jname}{E-mails are missing.\MessageBreak Plese use \protect\emailAdd\space macro to provide e-mails.}
\fi
\afterEmailSpace
\if!\@xtum!\else\noindent{\@xtum}\afterXtumSpace\fi
\if!\@abstract!\else\noindent{\renewcommand\baselinestretch{.9}\textsc{Abstract:}}\ \@abstract\afterAbstractSpace\fi
\if!\@keywords!\else\noindent{\textsc{Keywords:}} \@keywords\afterKeywordsSpace\fi
\if!\@arxivnumber!\else\noindent{\textsc{ArXiv ePrint:}} \href{http://arxiv.org/abs/\@arxivnumber}{\@arxivnumber}\afterArxivSpace\fi
\if!\@dedicated!\else\vbox{\small\it\raggedleft\@dedicated}\afterDedicatedSpace\fi
\ifnotoc\else
\iftoccontinuous\else\newpage\fi
\beforetochook\hrule
\tableofcontents
\afterTocSpace
\hrule
\afterTocRuleSpace
\fi
\setcounter{footnote}{0}
\pagestyle{myplain}\pagenumbering{arabic}
} 

\renewcommand{\baselinestretch}{1.1}\normalsize
\divide\@tempdima\baselineskip
\@tempcnta=\@tempdima
\skip\@mpfootins = \skip\footins
\renewcommand{\@dotsep}{10000}

\newcommand\ps@myplain{
\pagenumbering{arabic}
\renewcommand\@oddfoot{\hfill-- \thepage\ --\hfill}
\renewcommand\@oddhead{}}
\let\ps@plain=\ps@myplain

\newcommand\ps@titlepage{\renewcommand\@oddfoot{}\renewcommand\@oddhead{}}

\numberwithin{equation}{section}

\renewcommand\section{\@startsection{section}{1}{\z@}%
                                   {-3.5ex \@plus -1.3ex \@minus -.7ex}%
                                   {2.3ex \@plus.4ex \@minus .4ex}%
                                   {\normalfont\large\bfseries}}
\renewcommand\subsection{\@startsection{subsection}{2}{\z@}%
                                   {-2.3ex\@plus -1ex \@minus -.5ex}%
                                   {1.2ex \@plus .3ex \@minus .3ex}%
                                   {\normalfont\normalsize\bfseries}}
\renewcommand\subsubsection{\@startsection{subsubsection}{3}{\z@}%
                                   {-2.3ex\@plus -1ex \@minus -.5ex}%
                                   {1ex \@plus .2ex \@minus .2ex}%
                                   {\normalfont\normalsize\bfseries}}
\renewcommand\paragraph{\@startsection{paragraph}{4}{\z@}%
                                   {1.75ex \@plus1ex \@minus.2ex}%
                                   {-1em}%
                                   {\normalfont\normalsize\bfseries}}
\renewcommand\subparagraph{\@startsection{subparagraph}{5}{\parindent}%
                                   {1.75ex \@plus1ex \@minus .2ex}%
                                   {-1em}%
                                   {\normalfont\normalsize\bfseries}}

\def\fnum@figure{\textbf{\figurename\nobreakspace\thefigure}}
\def\fnum@table{\textbf{\tablename\nobreakspace\thetable}}

\long\def\@makecaption#1#2{%
  \vskip\abovecaptionskip
  \sbox\@tempboxa{\small #1. #2}%
  \ifdim \wd\@tempboxa >\hsize
    \small #1. #2\par
  \else
    \global \@minipagefalse
    \hb@xt@\hsize{\hfil\box\@tempboxa\hfil}%
  \fi
  \vskip\belowcaptionskip}

\renewenvironment{thebibliography}[1]{%
\begin{oldthebibliography}{#1}%
\small%
\raggedright%
\setlength{\itemsep}{5pt plus 0.2ex minus 0.05ex}%
}%
{%
\end{oldthebibliography}%
}

\begin{document}


\title{Scattering theory without large-distance asymptotics: scattering boundary condition}

\author[a]{Wen-Du Li}
\author[a,b,1]{and Wu-Sheng Dai}\note{daiwusheng@tju.edu.cn.}


\affiliation[a]{Department of Physics, Tianjin University, Tianjin 300072, P.R. China}
\affiliation[b]{LiuHui Center for Applied Mathematics, Nankai University \& Tianjin University, Tianjin 300072, P.R. China}








\abstract{By large-distance asymptotics, in conventional scattering theory,\ at the cost
of losing the information of the distance between target and observer, one
arrives at an explicit expression for scattering wave functions represented by
a scattering phase shift. In the present paper, together with a preceding
paper (T. Liu,W.-D. Li, and W.-S. Dai, JHEP06(2014)087), we establish a
rigorous scattering theory without imposing large-distance asymptotics. We
show that even without large-distance asymptotics, one can also obtain an
explicit scattering wave function represented also by a scattering phase
shift, in which, of course, the information of the distance is preserved.
Nevertheless, the scattering amplitude obtained in the preceding paper depends
not only on the scattering angle but also on the distance between target and
observer. In this paper, by constructing a scattering boundary condition
without large-distance asymptotics, we introduce a scattering amplitude, like
that in conventional scattering theory, depending only on the scattering angle
and being independent of the distance. Such a scattering amplitude, when
taking large-distance asymptotics, will recover the\ scattering amplitude in
conventional scattering theory. The present paper, with the preceding paper,
provides a complete scattering theory without large-distance asymptotics.}

\maketitle
\flushbottom


\section{Introduction}

In our preceding work, ref. \cite{liu2014scattering}, we establish a rigorous
scattering theory without imposing large-distance asymptotics. The information
of the distance between target and observer, which is lost in\ conventional
scattering theory due to large-distance asymptotics, is taken into account.
The scattering amplitude without large-distance asymptotics introduced in ref.
\cite{liu2014scattering}, however, depends both on scattering angle and
distance between target and observer. As a comparison, recall that in
conventional scattering theory, the scattering amplitude depends only on the
scattering angle.

In this paper, we present a scattering boundary condition without
large-distance asymptotics. Under this scattering boundary condition,
the\ scattering amplitude, as same as that in conventional scattering theory,
depends only on the scattering angle. This scattering boundary condition will
reduce to the Sommerfeld radiation condition, the scattering boundary
condition in conventional scattering theory, when taking large-distance asymptotics.

\textit{Scattering theory. }In quantum mechanics, all is determined by the
Schr\"{o}dinger equation with a given boundary condition. For bound-state
problems, the boundary condition is chosen to be $\left.  \psi\left(
\mathbf{r}\right)  \right\vert _{\mathbf{r}\in\text{boundary}}=0$, i.e., the
wave function vanishes on the boundary. For scattering problems, the boundary
condition is chosen to be a given wave function at an asymptotic distance.

Concretely, for a spherical potential, a scattering problem is determined by
the radial wave equation,%
\begin{equation}
\frac{1}{r^{2}}\frac{d}{dr}\left(  r^{2}\frac{dR_{l}}{dr}\right)  +\left[
k^{2}-\frac{l\left(  l+1\right)  }{r^{2}}-V\left(  r\right)  \right]  R_{l}=0,
\label{jingxiangeq}%
\end{equation}
with the scattering boundary condition,%
\begin{equation}
\psi=e^{ikr\cos\theta}+\psi^{sc}, \label{sbc-gen}%
\end{equation}
where $e^{ikr\cos\theta}$ is the incident plane wave and $\psi^{sc}$ is a
scattering wave function at an asymptotic distance.

To chose a scattering boundary condition is just to chose a $\psi^{sc}$. The
scattering amplitude is defined by $\psi^{sc}$, the scattering part of the
scattering boundary condition (\ref{sbc-gen}). Different choices of $\psi
^{sc}$ define different scattering amplitudes.

The starting of a scattering theory is the following result: The radial wave
equation (\ref{jingxiangeq})\ with $V\left(  r\right)  =0$ can be exactly
solved and the incident plane wave $e^{ikr\cos\theta}$ can be exactly
expanded:%
\begin{align}
R_{l}\left(  r\right)   &  =C_{l}h_{l}^{\left(  2\right)  }\left(  kr\right)
+D_{l}h_{l}^{\left(  1\right)  }\left(  kr\right)  ,\label{asyRl}\\
e^{ikr\cos\theta}  &  =\sum_{l=0}^{\infty}\left(  2l+1\right)  i^{l}%
j_{l}\left(  kr\right)  P_{l}\left(  \cos\theta\right)  , \label{pmbexpand}%
\end{align}
where $h_{l}^{\left(  1\right)  }\left(  z\right)  $ and $h_{l}^{\left(
2\right)  }\left(  z\right)  $ are the first and second kind spherical Hankel
functions, $j_{l}\left(  z\right)  $ the spherical Bessel function, and
$P_{l}\left(  x\right)  $ the Legendre polynomial \cite{olver2010nist}.

\textit{Conventional scattering theory}. In conventional scattering theory,
the scattering boundary condition is chosen as the Sommerfeld radiation
condition:
\begin{equation}
\psi=e^{ikr\cos\theta}+f\left(  \theta\right)  \frac{e^{ikr}}{r},\text{
\ \ }r\rightarrow\infty. \label{SRtiaojian}%
\end{equation}
$\psi^{sc}$ here is chosen as $\psi^{sc}=f\left(  \theta\right)  e^{ikr}/r$;
the scattering amplitude defined here is $f\left(  \theta\right)  $.

In coordination with the Sommerfeld radiation condition (\ref{SRtiaojian}),
the exact solution, eq. (\ref{asyRl}), and the exact expansion of the incident
plane wave, eq. (\ref{pmbexpand}), are approximately replaced by their
asymptotics \cite{ballentine1998quantum}:
\begin{align}
&  R_{l}\left(  r\right)  \overset{r\rightarrow\infty}{\sim}A_{l}\frac
{\sin\left(  kr-l\pi/2+\delta_{l}\right)  }{kr}, \label{yuanchangjingxiangjie}%
\\
&  e^{ikr\cos\theta}\overset{r\rightarrow\infty}{\sim}\sum_{l=0}^{\infty
}\left(  2l+1\right)  i^{l}\frac{\sin\left(  kr-l\pi/2\right)  }{kr}%
P_{l}\left(  \cos\theta\right)  , \label{ycpmb}%
\end{align}
where $\delta_{l}$ is the scattering phase shift.

The reason why in the Sommerfeld radiation condition (\ref{SRtiaojian})
$\psi^{sc}$ is chosen as being in proportion to $e^{ikr}/r$ is that the
asymptotic solution of the radial wave equation (\ref{jingxiangeq}) is
$R_{l}\overset{r\rightarrow\infty}{\sim}e^{\pm ikr}/r$ and only the outgoing
wave $R_{l}\overset{r\rightarrow\infty}{\sim}e^{ikr}/r$ remains in the
scattering wave function when $r\rightarrow\infty$ \cite{joachain1975quantum}.

The reason why the approximations (\ref{yuanchangjingxiangjie}) and
(\ref{ycpmb}) are used in conventional scattering theory is that only then can
the scattering phase shift $\delta_{l}$ appear explicitly.

\textit{Scattering theory without large-distance asymptotics}. In ref.
\cite{liu2014scattering}, we show that, even without large-distance
asymptotics, (\ref{yuanchangjingxiangjie}) and (\ref{ycpmb}), the two
approximations used in conventional scattering theory, one can still obtain a
scattering theory in which the phase shift can also appear explicitly, and, of
course, more information is preserved due to the fact that there is no approximation.

In ref. \cite{liu2014scattering}, the scattering boundary condition is taken
to be
\begin{equation}
\psi=e^{ik\cos\theta}+f\left(  r,\theta\right)  \frac{e^{ikr}}{r}.\label{jcSC}%
\end{equation}
$\psi^{sc}$ here is chosen as $\psi^{sc}=f\left(  r,\theta\right)  e^{ikr}/r$;
the scattering amplitude defined here is $f\left(  r,\theta\right)  $.

Moreover, the exact solution, eq. (\ref{asyRl}), and the expansion of the
incident plane wave, eq. (\ref{pmbexpand}), without any approximation, are
rewritten as \cite{liu2014scattering}%
\begin{align}
R_{l}\left(  r\right)   &  =M_{l}\left(  -\frac{1}{ikr}\right)  \frac{A_{l}%
}{kr}\sin\left[  kr-\frac{l\pi}{2}+\delta_{l}+\Delta_{l}\left(  -\frac{1}%
{ikr}\right)  \right]  ,\label{Rlr}\\
e^{ikr\cos\theta}  &  =\sum_{l=0}^{\infty}\left(  2l+1\right)  i^{l}%
M_{l}\left(  -\frac{1}{ikr}\right)  \frac{1}{kr}\sin\left[  kr-\frac{l\pi}%
{2}+\Delta_{l}\left(  -\frac{1}{ikr}\right)  \right]  P_{l}\left(  \cos
\theta\right)  ,
\end{align}
where $M_{l}\left(  x\right)  =\left\vert y_{l}\left(  x\right)  \right\vert $
and $\Delta_{l}\left(  x\right)  =\arg y_{l}\left(  x\right)  $ are the
modulus and argument of the Bessel polynomial $y_{l}\left(  x\right)  $, respectively.

It can be directly seen that the phase shift $\delta_{l}$ appears explicitly
in eq. (\ref{Rlr}), like that in conventional scattering theory, eq.
(\ref{yuanchangjingxiangjie}).

Nevertheless, the scattering boundary condition (\ref{jcSC}) is a naive
generalization of the Sommerfeld radiation condition. As a result, the
scattering amplitude defined by the scattering boundary condition
(\ref{jcSC}), $f\left(  r,\theta\right)  $, depends not only on the scattering
angle\ $\theta$ but also on the distance $r$, rather than the scattering
amplitude defined by the Sommerfeld radiation condition, $f\left(
\theta\right)  $. This implies that the scattering boundary condition
(\ref{jcSC}) is not very suitable for the scattering theory without
large-distance asymptotics, since the scattering part in eq. (\ref{jcSC}) is
in proportion to $e^{ikr}/r$ which is not an exact solution of the radial wave
equation (\ref{jingxiangeq}) with $V\left(  r\right)  =0$, unless
$r\rightarrow\infty$.

The aim of the present paper is, by constructing a
without-large-distance-asymptotics scattering boundary condition, to introduce
a scattering amplitude depending only on the scattering angle\ $\theta$ but
being independent of the distance $r$, like that in conventional scattering theory.

Without large-distance asymptotics, a new scattering boundary condition is
constructed in Sec. \ref{boundarycondition}. The expression of the scattering
amplitude defined by the new scattering boundary condition is also given in
this section. The relations among different scattering amplitudes, defined by
different scattering boundary conditions, are given in Sec. \ref{relations}.
The scattering cross section is considered in Sec. \ref{crosssection}. A
discussion on the scattering phase shift is given in \ref{Phase shift}.
Conclusions and outlook are given in Sec. \ref{conclusion}.

\section{Scattering boundary condition without large-distance asymptotics
\label{boundarycondition}}

In this section, 1) a scattering boundary condition without large-distance
asymptotics is presented, 2) by which a scattering amplitude depending only
on\ the scattering angle\ $\theta$ is defined.

In a word, in the following, we construct a \textit{distance-independent}
scattering amplitude without the help of the large-distance asymptotic
approximation. While, in ref. \cite{liu2014scattering}, we have to introduce a
\textit{distance-dependent} scattering amplitude for constructing a scattering
theory without large-distance asymptotics.

\subsection{Scattering boundary condition}

The reason why in conventional scattering theory the information of the
distance $r$ is lost is that the exact outgoing solution, $h_{l}^{\left(
1\right)  }\left(  kr\right)  $, is approximately replaced by its
large-distance asymptotics: $h_{l}^{\left(  1\right)  }\left(  kr\right)
\overset{r\rightarrow\infty}{\sim}e^{ikr}/r$. To retrieve the information of
the distance, without large-distance asymptotics, we construct a scattering
boundary condition by $h_{l}^{\left(  1\right)  }\left(  kr\right)  $ rather
than its asymptotics:%
\begin{equation}
\psi\left(  r,\theta\right)  =e^{ikr\cos\theta}+\sum_{l=0}^{\infty}%
a_{l}\left(  \theta\right)  h_{l}^{\left(  1\right)  }\left(  kr\right)
.\label{sanshebiantiaojian}%
\end{equation}
$\psi^{sc}$ here is chosen as $\psi^{sc}=\sum_{l=0}^{\infty}a_{l}\left(
\theta\right)  h_{l}^{\left(  1\right)  }\left(  kr\right)  $; the scattering
amplitude defined here is $a_{l}\left(  \theta\right)  $ which is a partial
wave scattering amplitude, rather than that in conventional scattering theory
and in ref. \cite{liu2014scattering}.

It can be directly verified that the scattering boundary condition
(\ref{sanshebiantiaojian}) reduces to the Sommerfeld radiation condition
(\ref{SRtiaojian}) when $r\rightarrow\infty$:%
\begin{equation}
\sum_{l=0}^{\infty}a_{l}\left(  \theta\right)  h_{l}^{\left(  1\right)
}\left(  kr\right)  \overset{r\rightarrow\infty}{\sim}f\left(  \theta\right)
\frac{e^{ikr}}{r}. \label{yuanchang}%
\end{equation}
The relation between $a_{l}\left(  \theta\right)  $ and $f\left(
\theta\right)  $ will be given below.

\subsection{Scattering amplitude $a_{l}\left(  \theta\right)  $
\label{amplitude}}

In this section, we calculate the scattering amplitude $a_{l}\left(
\theta\right)  $ defined by the scattering boundary condition
(\ref{sanshebiantiaojian}).

First, rewrite the scattering boundary condition (\ref{sanshebiantiaojian})
as
\begin{align}
\psi\left(  r,\theta\right)   &  =\sum_{l=0}^{\infty}\left(  2l+1\right)
i^{l}\frac{1}{2}\left[  h_{l}^{\left(  1\right)  }\left(  kr\right)
+h_{l}^{\left(  2\right)  }\left(  kr\right)  \right]  P_{l}\left(  \cos
\theta\right)  +\sum_{l=0}^{\infty}a_{l}\left(  \theta\right)  h_{l}^{\left(
1\right)  }\left(  kr\right) \nonumber\\
&  =\sum_{l=0}^{\infty}\left\{  \left(  2l+1\right)  i^{l}\frac{1}{2}%
P_{l}\left(  \cos\theta\right)  h_{l}^{\left(  2\right)  }\left(  kr\right)
+\left[  \left(  2l+1\right)  i^{l}\frac{1}{2}P_{l}\left(  \cos\theta\right)
+a_{l}\left(  \theta\right)  \right]  h_{l}^{\left(  1\right)  }\left(
kr\right)  \right\}  , \label{psi1}%
\end{align}
by use of the plane wave expansion (\ref{pmbexpand}) and $j_{l}\left(
x\right)  =\frac{1}{2}\left[  h_{l}^{\left(  1\right)  }\left(  x\right)
+h_{l}^{\left(  2\right)  }\left(  x\right)  \right]  $
\cite{liu2014scattering}.

Second, rewrite the radial wave function (\ref{asyRl}), the exact solution of
the radial wave equation (\ref{jingxiangeq}) with $V\left(  r\right)  =0$, as
\begin{equation}
R_{l}\left(  r\right)  =C_{l}h_{l}^{\left(  2\right)  }\left(  kr\right)
+D_{l}h_{l}^{\left(  1\right)  }\left(  kr\right)  =C_{l}\left[
h_{l}^{\left(  2\right)  }\left(  kr\right)  +e^{2i\delta_{l}}h_{l}^{\left(
1\right)  }\left(  kr\right)  \right]  ,
\end{equation}
where $e^{2i\delta_{l}}=D_{l}/C_{l}$ defines the phase shift
\cite{liu2014scattering}. The wave function $\psi\left(  r,\theta\right)
=\sum_{l=0}^{\infty}R_{l}\left(  r\right)  P_{l}\left(  \cos\theta\right)  $,
then, becomes%
\begin{equation}
\psi\left(  r,\theta\right)  =\sum_{l=0}^{\infty}\left[  C_{l}P_{l}\left(
\cos\theta\right)  h_{l}^{\left(  2\right)  }\left(  kr\right)  +C_{l}%
e^{2i\delta_{l}}P_{l}\left(  \cos\theta\right)  h_{l}^{\left(  1\right)
}\left(  kr\right)  \right]  . \label{psi2}%
\end{equation}

Finally, the scattering amplitude $a_{l}\left(  \theta\right)  $ can be
achieved immediately by equating the coefficients of $h_{l}^{\left(  1\right)
}\left(  kr\right)  $ and $h_{l}^{\left(  2\right)  }\left(  kr\right)  $ in
eqs. (\ref{psi1}) and (\ref{psi2}):
\begin{align}
C_{l}  &  =\left(  2l+1\right)  i^{l}\frac{1}{2},\label{Cl}\\
C_{l}e^{2i\delta_{l}}P_{l}\left(  \cos\theta\right)   &  =\left(  2l+1\right)
i^{l}\frac{1}{2}P_{l}\left(  \cos\theta\right)  +a_{l}\left(  \theta\right)  .
\label{alwithCl}%
\end{align}
Substituting $C_{l}$ into eq. (\ref{alwithCl}) gives
\begin{equation}
a_{l}\left(  \theta\right)  =\left(  2l+1\right)  i^{l}\frac{1}{2}\left(
e^{2i\delta_{l}}-1\right)  P_{l}\left(  \cos\theta\right)  . \label{a1}%
\end{equation}

Notice that the scattering amplitude $a_{l}\left(  \theta\right)  $ is
independent of the distance $r$.

\section{Relations among scattering amplitudes: $a_{l}\left(  \theta\right)
$, $f\left(  r,\theta\right)  $, and $f\left(  \theta\right)  $
\label{relations}}

A scattering process is fully described by the scattering amplitude. In
conventional scattering theory, \textit{with} large-distance asymptotics, the
scattering amplitude $f\left(  \theta\right)  $ is defined by the Sommerfeld
radiation condition (\ref{SRtiaojian}). In ref. \cite{liu2014scattering},
\textit{without} large-distance asymptotics, the scattering amplitude
$f\left(  r,\theta\right)  $, a function of the distance $r$, is defined by
the scattering boundary condition (\ref{jcSC}). In the present paper, also
\textit{without} large-distance asymptotics, the scattering amplitude
$a_{l}\left(  \theta\right)  $, which is independent of the distance $r$, is
defined by the scattering boundary condition (\ref{sanshebiantiaojian}).

In the following, we reveal the relations among these three scattering
amplitudes, $a_{l}\left(  \theta\right)  $, $f\left(  r,\theta\right)  $, and
$f\left(  \theta\right)  $.

\subsubsection{Relation between $f\left(  \theta\right)  $ and $a_{l}\left(
\theta\right)  $}

The relation between $a_{l}\left(  \theta\right)  $ and $f\left(
\theta\right)  $ can be achieved directly by performing large-distance
asymptotics in the scattering part of the scattering boundary condition
(\ref{sanshebiantiaojian}):%
\begin{equation}
\left.  \sum_{l=0}^{\infty}a_{l}\left(  \theta\right)  h_{l}^{\left(
1\right)  }\left(  kr\right)  \right\vert _{r\rightarrow\infty}=\left[
\sum_{l=0}^{\infty}a_{l}\left(  \theta\right)  \frac{1}{i^{l+1}k}\right]
\frac{e^{ikr}}{r}=f\left(  \theta\right)  \frac{e^{ikr}}{r},
\end{equation}
where the asymptotics of the spherical Hankel function, $h_{l}^{\left(
1\right)  }\left(  z\right)  \overset{r\rightarrow\infty}{\sim}\left(
-i\right)  ^{l+1}e^{iz}/z$ \cite{liu2014scattering}, is used.

Then we obtain the relation between $f\left(  \theta\right)  $ and
$a_{l}\left(  \theta\right)  $:
\begin{equation}
f\left(  \theta\right)  =\sum_{l=0}^{\infty}a_{l}\left(  \theta\right)
\frac{1}{i^{l+1}k}.
\end{equation}
It can be seen that the information of the distance $r$ loses when taking
large-distance asymptotics.

\subsubsection{Relation between $f\left(  r,\theta\right)  $ and $a_{l}\left(
\theta\right)  $}

Without large-distance asymptotics, we have defined two scattering amplitudes:
$f\left(  r,\theta\right)  $ and $a_{l}\left(  \theta\right)  $, where
$f\left(  r,\theta\right)  $ is defined by the scattering boundary condition
(\ref{jcSC}) given in ref. \cite{liu2014scattering} and $a_{l}\left(
\theta\right)  $ is defined by the scattering boundary condition
(\ref{sanshebiantiaojian}) given in the present paper. The relation between
$f\left(  r,\theta\right)  $ and $a_{l}\left(  \theta\right)  $ can be
achieved by comparing eqs. (\ref{jcSC}) and (\ref{sanshebiantiaojian}) directly:%

\begin{align}
f\left(  r,\theta\right)   &  =e^{-ikr}r\sum_{l=0}^{\infty}a_{l}\left(
\theta\right)  h_{l}^{\left(  1\right)  }\left(  kr\right) \nonumber\\
&  =\frac{1}{k}\sum_{l=0}^{\infty}a_{l}\left(  \theta\right)  \left(
-i\right)  ^{l+1}y_{l}\left(  -\frac{1}{ikr}\right)  , \label{frthtatht}%
\end{align}
where $h_{l}^{\left(  1\right)  }\left(  z\right)  =\left(  -i\right)
^{l+1}\left(  e^{iz}/z\right)  y_{l}\left(  i/z\right)  $
\cite{liu2014scattering} is used. From this result we can directly see how
does $f\left(  r,\theta\right)  $ depend on the distance $r$.

The scattering amplitude $f\left(  r,\theta\right)  $, introduced in ref.
\cite{liu2014scattering}, then, by the relation (\ref{frthtatht}), can be
obtained directly:
\begin{equation}
f\left(  r,\theta\right)  =\frac{1}{2ik}\sum_{l=0}^{\infty}\left(
2l+1\right)  \left(  e^{2i\delta_{l}}-1\right)  P_{l}\left(  \cos
\theta\right)  y_{l}\left(  -\frac{1}{ikr}\right)  ,
\end{equation}
which agrees with the result given in ref. \cite{liu2014scattering}.

\section{Scattering cross section \label{crosssection}}

In this section, we express the\ differential and total scattering cross
sections by the scattering amplitude introduced in the present paper,
$a_{l}\left(  \theta\right)  $.

\subsection{Differential scattering cross section}

Without large-distance asymptotics, in ref. \cite{liu2014scattering}, we
provide an exact expression of the differential scattering cross section. In
this section, for simplicity, we consider the leading contribution of the
differential scattering cross section,
\begin{equation}
\frac{d\sigma}{d\Omega}=\frac{\mathbf{j}^{sc}\cdot d\mathbf{S}}{j^{in}}%
=\frac{1}{k}\operatorname{Im}\left(  \psi^{sc\ast}\frac{\partial}{\partial
r}\psi^{sc}\right)  r^{2}.\label{diffscs}%
\end{equation}
Substituting $\psi^{sc}=\psi-\psi^{in}=\sum_{l=0}^{\infty}a_{l}\left(
\theta\right)  h_{l}^{\left(  1\right)  }\left(  kr\right)  $ (see eq.
(\ref{sanshebiantiaojian})) into eq. (\ref{diffscs}) and dropping the
high-order contribution, we achieve a differential scattering cross section
represented by the scattering amplitude $a_{l}\left(  \theta\right)  $,
\begin{equation}
\frac{d\sigma}{d\Omega}=r^{2}\left\vert \sum_{l=0}a_{l}\left(  \theta\right)
h_{l}^{\left(  1\right)  }\left(  kr\right)  \right\vert ^{2}.
\end{equation}

\subsection{Total scattering cross section}

The total scattering cross section can be achieved immediately by integrating
the differential scattering cross section. For simplicity, we only take the
leading contribution into account. We have
\begin{align}
\sigma\left(  r\right)   &  =4\pi r^{2}\sum_{l=0}\left(  2l+1\right)  \sin
^{2}\delta_{l}\left\vert h_{l}^{\left(  1\right)  }\left(  kr\right)
\right\vert ^{2}\nonumber\\
&  =\frac{4\pi}{k^{2}}\sum_{l=0}\left(  2l+1\right)  \sin^{2}\delta
_{l}\left\vert y_{l}\left(  -\frac{1}{ikr}\right)  \right\vert ^{2},
\end{align}
which agrees with the result given in ref. \cite{liu2014scattering}.

\section{A note on phase shift \label{Phase shift}}

In conventional scattering theory, with the help of large-distance
asymptotics, it is proved that the phase shift is the only effect in an
elastic scattering process, i.e., all information of an elastic scattering
process is embedded in a scattering phase shift \cite{ballentine1998quantum}.

In this section, we show that such a statement also holds without
large-distance asymptotics.

Without large-distance asymptotics, as shown in (\ref{psi1}), the incident
plane wave can be exactly expressed as
\begin{equation}
\psi^{in}\left(  r,\theta\right)  =\sum_{l=0}^{\infty}\frac{1}{2}\left(
2l+1\right)  i^{l}\left[  h_{l}^{\left(  2\right)  }\left(  x\right)
+h_{l}^{\left(  1\right)  }\left(  x\right)  \right]  P_{l}\left(  \cos
\theta\right)  . \label{psiin}%
\end{equation}
After an elastic scattering, by eq. (\ref{psi2}), the wave function becomes
(for clarity and convenience, we rewrite eq. (\ref{psi2}) here)%
\begin{equation}
\psi\left(  r,\theta\right)  =\sum_{l=0}^{\infty}\frac{1}{2}\left(
2l+1\right)  i^{l}\left[  h_{l}^{\left(  2\right)  }\left(  kr\right)
+e^{2i\delta_{l}}h_{l}^{\left(  1\right)  }\left(  kr\right)  \right]
P_{l}\left(  \cos\theta\right)  . \label{psi2copy}%
\end{equation}

Comparing the wave functions before and after the scattering, eqs.
(\ref{psiin}) and (\ref{psi2copy}), we can see that, the incoming part,
represented by $h_{l}^{\left(  2\right)  }\left(  kr\right)  $, does not
change anymore, while a phase factor $e^{2i\delta_{l}}$ appears in the
outgoing part, represented by $h_{l}^{\left(  1\right)  }\left(  kr\right)  $.
This shows that the only effect after an elastic scattering is a phase shift
on the outgoing wave function, even without large-distance asymptotics.

\section{Conclusions and outlook \label{conclusion}}

A scattering theory without using large-distance-asymptotics approximation is
established in our two papers, ref. \cite{liu2014scattering} and the present
paper: in ref. \cite{liu2014scattering}, we deal with the solution of the
radial wave equation and the incident wave; in the present paper, we deal with
the scattering boundary condition. Now, we have a complete scattering theory
without large-distance asymptotics and, thus, without losing the information
of distance.

Concretely, a scattering boundary condition without large-distance asymptotics
is constructed in this paper. The scattering amplitude defined by such a
scattering boundary condition, $a_{l}\left(  \theta\right)  $, rather than
$f\left(  r,\theta\right)  $ (the scattering amplitude introduced in ref.
\cite{liu2014scattering}), depends only on the scattering angle $\theta$ and
is independent of the distance $r$.

In further works, based on the scattering theory without large-distance
asymptotics, we will systematically deal with a series of scattering related
problems which are all treated under large-distance asymptotics in the frame
of conventional scattering theory. Without large-distance asymptotics, we can
also construct a complete treatment on the Lippmann-Schwinger equation. In
conventional scattering theory, a very important problem is the analytic
property of scattering amplitudes, which, of course, based on large-distance
asymptotics
\cite{arnecke2008jost,willner2006low,laha2005off,decanini2003complex}. Such a
problem, now, can be discussed without large-distance asymptotics. Based on
two important quantum field theory methods, scattering spectrum method
\cite{rahi2009scattering,forrow2012variable,bimonte2012exact} and heat kernel
method
\cite{vassilevich2003heat,fucci2012heat,moral2012derivative,dai2010approach},
we establish a heat-kernel method for the calculation of the scattering phase
shift through the relation between scattering spectrum method and heat kernel
method given by Refs. \cite{dai2009number,pang2012relation}; now, we can do
this without large-distance asymptotics. The scattering theory given by ref.
\cite{liu2014scattering}) and the present paper is in fact a scalar scattering
theory. Therefore, the result can be applied to any scalar scattering, such as
acoustic scatterings. In an acoustic scattering, in comparison with the
distance between target and observer, the wave length of an acoustic wave
often cannot be ignored; in such cases, one can use our result to construct an
acoustic scattering\ theory without imposing large-distance asymptotics, while
in conventional acoustic scattering theory, large-distance asymptotics
(far-field pattern) is imposed
\cite{berthet2003using,ikehata2012inverse,liu2013near}. Moreover, the scalar
scattering theory without large-distance asymptotics can be naturally
generalized to vector and tensor scatterings. The scattering of
electromagnetic waves is a vector scattering. We can establish a scattering
theory of electromagnetic waves without large-distance asymptotics, especially
for long wavelength cases. By generalizing our result to tensor wave
scattering, we can consider the scattering theory of gravitational waves,
which has been studied under large-distance asymptotics in literature, e.g.,
\cite{dolan2008scattering,giddings2010gravitational}. Moreover, we can also
consider the scattering of a wave scattered by a black hole; all discussions
on this issue are based on large-distance asymptotics
\cite{crispino2009electromagnetic,doran2002perturbation,dolan2006fermion,crispino2009scattering,raffaelli2013scattering}%
. A relativistic scattering theory without large-distance asymptotics also can
be established. Inverse scattering problems
\cite{chadan1997introduction,sabatier2000past} can also be systematically
studied in the frame of the scattering theory without large-distance
asymptotics. More topics on scatterings can be found in ref.
\cite{pike2001scattering}. Moreover, many scattering theories based on
large-distance asymptotics can be treated without such an asymptotic
approximation \cite{costa2012conformal,chen2007quantum,hod2013scattering}.


\acknowledgments

We are very indebted to Dr G. Zeitrauman for his encouragement. This work is
supported in part by NSF of China under Grant No. 11075115.



\providecommand{\href}[2]{#2}\begingroup\raggedright\endgroup







\end{document}